# Reconstruction of electron velocity distribution function and Gibbs entropy from electron cyclotron emission in magnetized plasmas

Kawamori Eiichirou

Institute of Space and Plasma Sciences, National Cheng Kung University, Taiwan

E-mail: kawamori@isaps.ncku.edu.tw

**Abstract**

We propose a method for reconstructing the fluctuation components of the electron velocity distribution function $\delta f(v_\perp)$, and the electron entropy, which is a functional of $\delta f(v_\perp)$ expressed as $-\int \delta f(v_\perp) \ln \delta f(v_\perp) dv_\perp$, using the harmonic spectrum from pure X-mode electron cyclotron emission (ECE) in optically thin plasmas. Here, $v_\perp$ represents the electron velocity perpendicular to the background magnetic field. This formulation employs the maximum entropy method in velocity space using the Hankel transform, which converts from $v_\perp$ space to $p$ space (where $p$ is the index of the wavenumber in velocity space). Numerical tests validated the effectiveness of the proposed method, which is applicable across a wide range of magnetized plasma conditions, including conditions with both non-relativistic and relativistic electrons, except in cases of harmonic overlap or under optically thick conditions. Notably, this method does not require radiometer calibration for ECE measurements. This method fascilitates the experimental evaluation of electron entropy transport in fusion plasma experiments. Moreover, when combined with measurements in $k$-space (spatial distribution), this approach enables entropy distribution acquisition in phase space ($k$-$p$ space).

## 1. Introduction

Understanding the irreversible processes involved in energy transport through magnetized plasma turbulence requires a thorough assessment of the entropy transport in phase space. Since the entropy is directly related to the particle velocity distribution function, accurately measuring the distribution function is essential for evaluating entropy.

The transport of entropy plays a critical role in deciphering the dynamics of magnetized plasma turbulence [1-9]. Measuring fluctuations in the particle velocity distribution functions within phase space is vital for understanding the entropy transport. For ions, an entropy cascade in turbulent phase space has been successfully observed in laboratory plasmas through direct measurements of velocity distribution fluctuations [10]. However, measuring fluctuations in the velocity distribution function is notably more challenging for electrons than for ions.

Hutchinson and Kato proposed a method for reconstructing the electron velocity distribution function (EVDF), $f(v)$, based on the harmonic spectrum of electron cyclotron emission (ECE) for mildly relativistic electrons [11,12,13]. Their approach leverages the relativistic frequency shift associated with ECE, which limits its applicability to plasmas containing a considerable fraction of mildly relativistic electrons.

In contrast, the proposed method is applicable to a wide range of magnetized plasma scenarios, including both non-relativistic and relativistic electrons, with the exception of cases in which harmonic overlap, optically thick conditions, or similar constraints occur.

In this study, we propose a method for measuring the fluctuation components of electron entropy—a functional of the fluctuation component of EVDF, $\delta f(v_\perp)$—expressed as $-\int \delta f(v_\perp)\ln \delta f(v_\perp)dv_\perp$—using the harmonic spectrum ECE. We also report on the verification of our method through numerical tests. To reconstruct the EVDF, we introduce the maximum entropy method (MEM) in velocity space via the Hankel transform (HT), which transforms from $v_\perp$ space to $p$ space, where $p$ is the wavenumber index in velocity space. One advantage of this method is that the entropy representation in $p$ space can be obtained simultaneously during the reconstruction of $\delta f(v_\perp)$. In addition to measurements in $k$-space (spatial distribution measurements), this approach allows us to derive the entropy distribution in phase space.

## 2. Formulation

This study focuses on pure X-mode propagation. The spectral emissivity, $\eta \equiv (d^2P)/(d\omega d\Omega_k)$, of pure X-mode ECE from optically thin plasmas is given by the following equation (Eq. (41) in Ref. 14, Eq. (4) in Ref. 11, etc.). Here, $P$ represents the power of the ECE measured within the corresponding frequency band, and $d\Omega_k$ is an element of the solid angle subtended by the wave vector $\boldsymbol{k}$.

$$\eta(\omega) = \frac{e^2 n_e}{2\pi} \frac{\omega^2}{c^2} \int d\vec{u} f(\boldsymbol{u})$$
$$\times \sum_i \int_0^\infty dn^2 \frac{n\lambda_{ss}(\boldsymbol{k},\omega)}{\left|\left(\frac{\partial}{\partial n^2}\right)\Lambda_r(\boldsymbol{k},\omega)\right|} \delta(n^2 - n_i^2)$$
$$\times \left|\vec{a}_i^*(\boldsymbol{k},\omega)\cdot\vec{j}_m\right|^2 \delta\left[\omega - m\frac{\Omega_{ce}}{\gamma}\right], \qquad (1)$$

where $n^2 = c^2k^2/\omega^2$ is the index of refraction. Here, $u = |\boldsymbol{u}|$ is the relativistic velocity ($\boldsymbol{u} = \boldsymbol{p}/m_e = \gamma\boldsymbol{v}$), where $\gamma = (1+[u^2/c^2])^{1/2}$. The term $\lambda_{ss}(\boldsymbol{k},\omega)$ denotes the trace of $\lambda_{ij}^\dagger(\boldsymbol{k},\omega)$, where $\sum_j \lambda_{ij}(\boldsymbol{k},\omega)E_j = 0$ provides the wave equation with $E_j$ as the wave electric field vector and $\lambda_{ij}^\dagger$ indicates the classical adjoint of $\lambda_{ij}$. The vector $\vec{a}_i(\boldsymbol{k},\omega)$ is the unit polarization vector for the propagating mode $i$, and $\Lambda_r(\boldsymbol{k},\omega) = 0$ gives the dispersion relation for these modes. The vector $\vec{j}_m$ is defined as:

$$\vec{j}_m = \left(\frac{m\Omega_{ce}v_\perp}{k_\perp u_\perp}J_m\left(\frac{k_\perp u_\perp}{\Omega_{ce}}\right), iv_\perp J_m'\left(\frac{k_\perp u_\perp}{\Omega_{ce}}\right), v_{//}J_m\left(\frac{k_\perp u_\perp}{\Omega_{ce}}\right)\right). \qquad (2)$$

Here, $J_m$ is the $m$-th order Bessel function of the first kind; $\Omega_{ce}$ is the electron cyclotron frequency, and $k_\perp$ is the wavenumber of the X-wave. The second line of Eq. (1) represents the propagator of waves in the plasma, driven by the single particle source current $\vec{j}_m$. The term "$\left|\vec{a}_i^*(\boldsymbol{k},\omega)\cdot\vec{j}_m\right|^2$" represents the coupling between $\vec{j}_m$ and the propagation modes (with $i$ indexing the propagation modes).

It is important to note that $\eta$ is proportional to the electron density $n_e$, unlike in the case of blackbody radiation.

### 2.1 Non-Relativistic Case

In this Subsection, our formulation is constrained in non-relativistic regimes, where $\gamma \approx 1$ and $u_\perp \approx v_\perp$. In addition, we focus on the fluctuation components of EVDF, which are represented as $\delta f$. The pure X-modes of the ECE are considered, as mentioned.

In Eq. (1), $n_e$ and $f(v)$ contain both the equilibrium and the fluctuation components. That is,
$$n_e = n_{e0} + \tilde{n}_e, \qquad (3)$$
$$f(\boldsymbol{v}) = F_0(\boldsymbol{v}) + \delta f(\boldsymbol{v}), \qquad (4)$$
where $\tilde{n}_e$ is the fluctuation of the electron density. The frequencies of $\tilde{n}_e$ and $\delta f$ are much smaller than that of $\Omega_{ce}$, which is the characteristic frequency of $\vec{j}_m$. The term $\mathfrak{P}_m \equiv \sum_i \int_0^\infty dn^2 \frac{n\lambda_{ss}(\boldsymbol{k},m\Omega_{ce})}{\left|\left(\frac{\partial}{\partial n^2}\right)\Lambda_r(\boldsymbol{k},m\Omega_{ce})\right|} \delta(n^2 - n_i^2)$ in Eq.(1), which represents the propagator of waves in the plasma, is assumed to be less sensitive to $\tilde{n}_e$ and $\delta f$. The propagator and polarization are determined by the bulk cold plasma, which does not cause significant absorption and does not require the dielectric tensor to be treated kinetically or in a relativistic way.

*2.1.1 Evaluation of Entropy Stemming from the Fluctuation of $\delta f(v)$* The electron entropy density that is associated with the $\delta f(v)$ component of the lowest order can be expressed as:

$$\tilde{S} \equiv -\int \delta f^2 d\boldsymbol{v}. \qquad (5)$$

Accordingly, the fluctuation of the spectral emissivity is expressed as:

$$\tilde{\eta}_m(m\Omega_{ce})$$
$$= \mathfrak{P}_m e^2 (n_{e0} + \tilde{n}_e)\frac{(m\Omega_{ce})^2}{c^2}\int dv_\perp v_\perp$$
$$[F_0(v_\perp) + \delta f(v_\perp)]\sum_i |\vec{a}_i^*(k_\perp, m\Omega_{ce})\cdot\vec{j}_m|^2 \delta\left[\omega - m\frac{\Omega_{ce}}{\gamma}\right],$$
$$(m = 1,2,\dots). \qquad (6)$$

where $\delta f(v_\perp) \equiv \int_{-\infty}^\infty dv_{//}\,\delta f(v_\perp, v_{//})$. Please note that the term $\mathfrak{P}_m$ is dependent on only $F_0$.

For illustrative purposes, $\vec{a}^*(\boldsymbol{k},\omega)$ is approximated by $(2^{-1/2}, -i2^{-1/2}, 0)$ (i.e., right-hand circular polarization $\vec{a}(\boldsymbol{k},\omega) = (2^{-1/2}, i2^{-1/2}, 0)$) in Eq. (6) as follows:

$$\tilde{\eta}_m(m\Omega_{ce}) = \mathfrak{P}_m e^2 n_e \frac{(m\Omega_{ce})^2}{c^2} \int_0^\infty v_\perp dv_\perp [F_0(v_\perp) + \delta f(v_\perp)] \left[ \left(\frac{m\Omega_{ce}}{k_\perp}\right)^2 J_m^2(k_\perp \rho_e) + (\rho_e \Omega_{ce})^2 J_m'^2(k_\perp \rho_e) + J_{m-1}(k_\perp \rho_e) J_{m+1}(k_\perp \rho_e) \right] \delta\left[\omega - m\frac{\Omega_{ce}}{\gamma}\right]. \quad (7)$$

where $\rho_e$ is the electron Larmor radius. The Bessel functions in $\vec{j}_m$ account for the finite Larmor radius effect, i.e., the dependence of the coupling between the electron motions expressed by $\vec{j}_m$ and the propagation modes on the electron velocity.

*2.1.2 Application of MEM and HT* The equilibrium component of the spectral emissivity is expressed as:

$$\eta_{m0}(m\Omega_{ce}) = \mathfrak{P}_m e^2 n_{e0} \frac{(m\Omega_{ce})^2}{c^2} \int dv_\perp v_\perp [F_0(v_\perp)] \sum_i |\vec{a}_i^*(k_\perp, m\Omega_{ce}) \cdot \vec{j}_m|^2 \delta\left[\omega - m\frac{\Omega_{ce}}{\gamma}\right], \quad (m = 1, 2, \dots). \quad (8)$$

where $F_0(v_\perp) \equiv \int_{-\infty}^\infty dv_{//} F_0(v_\perp, v_{//})$. The first order fluctuation component of the spectral emissivity can be obtained from Eq. (6) and Eq. (8) for *m*-th harmonics as follows:

$$\frac{\tilde{\eta}_m - \eta_{m0}}{\eta_{m0}} = \frac{\tilde{n}_e}{n_{e0}} + \frac{\int v_\perp dv_\perp \delta f(v_\perp) \sum_i |\vec{a}_i^*(k_\perp, m\Omega_{ce}) \cdot \vec{j}_m|^2}{\int v_\perp dv_\perp F_0(v_\perp) \sum_i |\vec{a}_i^*(k_\perp, m\Omega_{ce}) \cdot \vec{j}_m|^2}. \quad (9)$$

Based on the formulation in Eq. (9), obtaining $\delta f$ from the measurement of the ECE spectrum $\frac{\tilde{\eta}_m - \eta_{m0}}{\eta_{m0}}$ is posed as an inverse problem. However, Eq. (9) is the Fredholm integral equation of the first-kind, which is generally ill-posed. Under the definition for $\tilde{S}$ shown in Eq. (5), the target function $\delta f$ is reconstructed by finding the parameter $\lambda_m$ that maximizes $\tilde{S}$ using the Lagrange multipliers with the ECE power measurements as constraints given that $\tilde{n}_e/n_{e0}$ is measured, e.g., by a reflectometer and Thomson scattering, and assuming the Maxwellian form of $F_0(v_\perp)$ in calculating the denominator of the second term on the RHS, Eq. (9). $\delta f$ is determined by solving $\delta I = 0$ with respect to the functional $I$, which is defined as follows:

$$I \equiv \tilde{S} - \sum_{m=1,\dots} \lambda_m \times \left\{ \tilde{\eta}_m(m\Omega_{ce}) - \mathfrak{P}_m e^2 (n_{e0} + \tilde{n}_e) \frac{(m\Omega_{ce})^2}{c^2} \int dv_\perp v_\perp [F_0(v_\perp) + \delta f(v_\perp)] \sum_i |\vec{a}_i^*(k_\perp, m\Omega_{ce}) \cdot \vec{j}_m|^2 \delta\left[\omega - m\frac{\Omega_{ce}}{\gamma}\right] \right\}. \quad (10)$$

The constraints are given by Eq. (9). Please note that Eq. (9) is independent of the other harmonic components; this eliminates the need for calibration between the different harmonic components. Additionally, the propagator $\mathfrak{P}_m$ has been removed from the formulation. The vectors $\vec{a}_i^*(k_\perp, m\Omega_{ce})$ and $k_\perp(\omega = m\Omega_{ce})$ in $\vec{j}_m$ are determined by calculating the dispersion relations of their propagation modes using $B_0$ and $n_{e0}$, where $B_0$ and $T_{e0}$ are the background magnetic field strength and electron temperature, respectively.

Here, we apply the Hankel transform (HT) to $\delta f$. The relationship between $\delta f(v_\perp)$ and $\delta f_p$ is given by the HT and its inverse transform:

$$\delta f_p = \frac{2}{(v_\perp^{ub})^2 J_1^2(j_{0p})} \int_0^{v_\perp^{ub}} v_\perp dv_\perp J_0\left(j_{0p}\frac{v_\perp}{v_\perp^{ub}}\right) \delta f(v_\perp), \quad (11)$$

$$\delta f(v_\perp) = \sum_{p=1}^{p_{ub}} \delta f_p J_0\left(j_{0p}\frac{v_\perp}{v_\perp^{ub}}\right), \quad (12)$$

where $v_\perp^{ub}$ is the upper limit of the velocity component, $p_{ub}$ is the upper limit of the index $p$, and $j_{0p}$ is the *p*-th zero of the 0-th order Bessel function of the first kind. The entropy in *p* space is expressed as:

$$\tilde{S} \equiv -\left(v_\perp^{ub}\right)^2 \sum_{p=1}^{p_{ub}} \delta f_p^2 \frac{J_1^2(j_{0p})}{2}. \quad (13)$$

It should be noted that entropy is negative definite (see also Eq. (5)), because the equilibrium state corresponds to the maximum entropy state.

By substituting Eq. (12) into Eq. (10) and setting the variation of Eq. (10) $\delta I = 0$, we obtain the Hankel-transformed $\delta f$ in *p*-space:

$$\delta f_p = \frac{(v_\perp^{ub})^2}{2 J_1(j_{0p})} \sum_{m=1}^{m_{max}} \lambda_m H_p^m(k_\perp(m\Omega_{ce})), \quad (14)$$

$$H_p^m(k_\perp(m\Omega_{ce})) = \int_0^1 x dx [\sum_i |\vec{a}_i^*(k_\perp, m\Omega_{ce}) \cdot \vec{j}_m(k_\perp, v_\perp)|^2] J_0(j_{0p} x). \quad (15)$$

The function $H_p^m$ can be computed once the propagation mode $i$ is determined, i.e., by calculating $k_\perp$ and $\vec{a}_i^*(k_\perp, m\Omega_{ce})$. This function serves as a type of basis function. The constant $\lambda_m / \int v_\perp dv_\perp F_0(v_\perp) \sum_i |\vec{a}_i^*(k_\perp, m\Omega_{ce}) \cdot \vec{j}_m|^2$ can be treated as a redefined $\lambda_m$.

By substituting Eq. (12) into the constraints Eq. (9), we obtain simultaneous equations for λ$_m$ (where $m$ = 1, 2, …) in $p$-space as follows:

$$\frac{\tilde{\eta}_m - \eta_{m0}}{\eta_{m0}} - \frac{\tilde{n}_e}{n_{e0}} = \sum_{p=1}^{p_{max}} \frac{(v_\perp^{ub})^2}{2J_1(j_{0p})} \times$$
$$\left\{\sum_{m'=1}^{m'_{max}} \lambda_{m'} H_p^{m'}(k_\perp(m\Omega_{ce}))\right\} H_p^m(k_\perp(m\Omega_{ce})). \quad (16)$$

The values of λ$_m$ are determined by solving these simultaneous equations. Once the λ$_m$ values are determined, $\delta f_p$ can be obtained from Eq. (14).

*2.2 Relativistic Case*

When the relativistic frequency shift $\Delta\omega$ is appreciable—specifically when $\gamma$ for a dominant velocity component is greater than 1—it becomes possible to analyze the dependence of $\delta f(\mathbf{u})$ on $u_{//} = \gamma v_{//}$. The spectral emissivity at frequency $m(\Omega_{ce} - \Delta\omega)$ ($m$ = 1, 2, …) restricts the electron radiation source to the momentum sphere defined by $\bar{\bar{\gamma}} = \sqrt{1 + \bar{\bar{u}}^2/c^2} = m\frac{\Omega_{ce}}{\omega} = \frac{\Omega_{ce}}{(\Omega_{ce} - \Delta\omega)}$. In this case, after the integration in the $u_{//}$ direction, Eq. (1) simplifies to:

$$\eta[m(\Omega_{ce} - \Delta\omega)] = e^2 n_e \frac{\{m(\Omega_{ce} - \Delta\omega)\}^2}{c^2}$$
$$\times \sum_i \int_0^\infty dn^2 \frac{n\lambda_{ss}(\mathbf{k}, m(\Omega_{ce} - \Delta\omega))}{\left|\left(\frac{\partial}{\partial n^2}\right)\Lambda_r(\mathbf{k}, m(\Omega_{ce} - \Delta\omega))\right|}\delta(n^2 - n_i^2)$$
$$\times \int_0^{\bar{\bar{u}}} du_\perp u_\perp \delta\hat{f}(u_\perp, |u_{//}|) \left|\vec{a}_i^*(\mathbf{k}, m(\Omega_{ce} - \Delta\omega)) \cdot \vec{J}_m\left(u_\perp, \gamma = \frac{\Omega_{ce}}{\omega_0}\right)\right|^2 \frac{\Omega_{ce}^2 c^2}{m(\Omega_{ce} - \Delta\omega)^3 \bar{\bar{u}}},$$
where $\bar{\bar{u}} \equiv \sqrt{\left(\frac{\Omega_{ce}^2}{(\Omega_{ce} - \Delta\omega)^2} - 1\right)c^2}$ and $\delta\hat{f}(u_\perp, |u_{//}|) \equiv \frac{\bar{\bar{u}}}{\sqrt{\bar{\bar{u}}^2 - u_\perp^2}}\left\{\delta f\left(u_\perp, \sqrt{\bar{\bar{u}}^2 - u_\perp^2}\right) + \delta f\left(u_\perp, -\sqrt{\bar{\bar{u}}^2 - u_\perp^2}\right)\right\}. \quad (17)$

The measured frequency $m(\Omega_{ce} - \Delta\omega)$, ($m$ = 1, 2, …), defines the radius of a circle in the $u$-space, specifically $\bar{\bar{\gamma}} - 1$. This facilitates the application of the reconstruction scheme for $\delta f(v_\perp)$, as elaborated upon in the previous section (Section 2.1) to reconstruct $\delta\hat{f}(u_\perp, |u_{//}|)$ as presented in Eq. (17). Furthermore, this method enables the determination of the dependence of $\delta f(\mathbf{v}) = \delta f(u_\perp, |u_{//}|)$ on $|u_{//}|$ using the relationship $u_{//}^2 = \sqrt{\bar{\bar{u}}^2 - u_\perp^2}$. However, it is important to note that this method cannot distinguish between positive and negative $u_{//}$, as described in the Hutchinson–Kato papers [11−13]. To successfully apply this reconstruction method, it is crucial that the original electron cyclotron frequency $\Omega_{ce}$ is accurately known. One effective configuration to meet this requirement involves viewing the tokamak plasma vertically from the top of the machine, thereby isolating a region of constant magnetic field [12, 13]. In this configuration, the EVDF obtained through the MEM-HT is line-integrated.

### 3. Experimental Application

This section outlines the procedure for using the $\delta f$ reconstruction method in experiments in the non-relativistic case. The steps for implementing the MEM–HT scheme are as follows:

i) Fourier transform:
Transform the time-domain ECE spectral emissivity $\tilde{\eta}_m(t, m\Omega_{ce})$ to the frequency domain to obtain $\tilde{\eta}_m(\omega, m\Omega_{ce})$, where $\omega \ll \Omega_{ce}$.
ii) Determine $\lambda_m$:
Determine $\lambda_m$ as the lease-square solutions from Eq. (16). $k_\perp(m\Omega_{ce})$ is calculated from the dispersion relation using $B_0$, $T_{e0}$, and $n_{e0}$.
iii) Obtain $\delta f_p(\omega)$:
Calculate $\delta f_p(\omega)$ using the relation from Eq. (14).
iv) Calculate entropy:
Compute the entropy $\tilde{S}(\omega, p)$ using Eq. (13). The total can be calculated by the summation $\tilde{S}(p) = \sum_\omega \tilde{S}(\omega, p)$.

### 4. Numerical Verification and Discussion

The numerical test of the reconstruction scheme was carried out as follows: i) We prepared $\delta f$ or eqivalently $\delta f_p$ as the "given" $\delta f$ or $\delta f_p$. ii) Using Eq. (9), we calculate the ratios $\frac{\tilde{\eta}_m - \eta_{m0}}{\eta_{m0}}$ as measured quantities that range from $m = m_{min}$ to $m = m_{max}$, where $m_{min}$ and $m_{max}$ denote the minimum and maximum harmonic numbers, respectively. For simplicity, we assume that $\tilde{n}_e = 0$. iii) We apply the procedure of the MEM–HT method (Section 3) to obtain $\delta f^{Recon}$ and $\delta f_p^{Recon}$ as "reconstructed" $\delta f$ and $\delta f_p$. A cold plasma approximation is assumed for the propagator $\mathfrak{P}_m$. The right-hand circular polarization for $\vec{a}^*(\mathbf{k}, \omega)$, which can be represented as $(2^{-1/2}, -i2^{-1/2}, 0)$ $[\vec{a}(\mathbf{k}, \omega) = (2^{-1/2}, i2^{-1/2}, 0)]$ for all harmonic ECEs, is assumed for simplicity. A non-relativistic treatment $\gamma = 1$, such that $u_\perp \approx v_\perp$, is also used. Correspondingly, the basis function in (15) is simplified as follows:

$$H_p^m(k_\perp(m\Omega_{ce})) = (v_\perp^{ub})^2 \int_0^1 x dx B_m J_0(j_{0p} x). \quad (18)$$

$$B_m \equiv x^2 \left[ J_{m-1}^2\left(\frac{ck_\perp}{\Omega_{ce}} \frac{v_\perp^{ub}}{c} x\right) + J_{m+1}^2(\cdot) + J_{m-1}(\cdot) J_{m+1}(\cdot) \right]. \quad (19)$$

where "·" represents $\frac{ck_\perp}{\Omega_{ce}} \frac{v_\perp^{ub}}{c} x$.

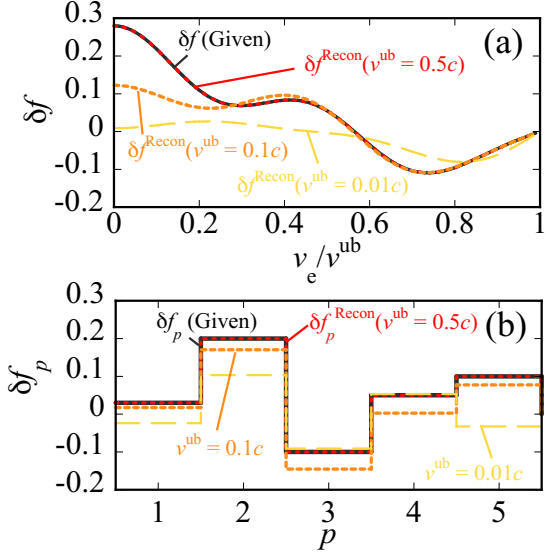

Figure 1. (a) The given $\delta f$ (black solid lines) and the reconstructed profiles for $v^{ub} = 0.01c$ (yellow dashed lines), $0.1c$ (orange dotted lines), and $0.5c$ (red dotted lines). The abscissa denotes the normalized electron velocity $v_\perp/v^{ub}$. (b) The corresponding $\delta f_p$, with the abscissa representing the mode number in $p$-space.

Figure 1 (a) illustrates the given $\delta f$ (black solid lines) and reconstructed profiles for $v^{ub} = 0.01c$ (yellow dashed lines), $0.1c$ (orange dotted lines), and $0.5c$ (red dotted lines). The distribution is given by:

$$\delta f(v_\perp) = \sum_{p=1}^{p_{ub}=5} \delta f_p J_0\left(j_{0p} \frac{v_\perp}{v_\perp^{ub}}\right),$$

where $\delta f_p = (0.03, 0.2, -0.1, 0.05, \text{and } 0.1)$. The abscissa represents the normalized electron velocity $v_\perp/v^{ub}$. The corresponding $\delta f_p$ is dispalyed in figure 1 (b), with the horizontal axis indicating the mode number $p$ in $p$-space. The reproducibility of $\delta f$ and $\delta f_p$ improves appreciably with increasing $v^{ub}$, and both are fully reconstructed at $v^{ub} = 0.5c$. Although the $p$ profile reproducibility is high across all $v^{ub}$ values, the reproducibility of the $v_\perp$ profile at $v^{ub} = 0.01c$ appears less effective. This suggests that the MEM–HT method extracts information in $p$-space as long as $p_{max}$ is limited by $m_{max}$. Please note that Eq. (1) remains well-posed (i.e., the matrix $H_p^m$ is diagonal) when $m_{min} = 0$ and $m_{max} = p_{max}$. Additionally, if the argument of $J_m$ and $J_m'$ in $H_p^m$, namely $\frac{k_\perp u_\perp}{\Omega_{ce}}$, is adequately large, the rank of $H_p^m$ approaches $m_{max} = p_{max}$.

This behavior arises from the variation in the $B_m$ profile relative to its argument scale in $H_p^m$ (Eq. (18)), particularly regarding $J_{m-1}^2\left(\frac{ck_\perp}{\Omega_{ce}} \frac{v_\perp^{ub}}{c} x\right)$ and $J_{m+1}^2\left(\frac{ck_\perp}{\Omega_{ce}} \frac{v_\perp^{ub}}{c} x\right)$. The function $B_m$ begins to increase from nearly zero when its argument is approximately $0.1m$ (this increasing point is slightly dependent on $m$). However, since the argument of $B_m$ is composed of three elements: $\frac{ck_\perp}{\Omega_{ce}} \sim m$, $0 \leq x \leq 1$, and $\frac{v_\perp^{ub}}{c}$ (bounded by 1), when $v_\perp^{ub}$ is much smaller than 1, $B_m$ does not develop within the range $0 \leq x \leq 1$ for any order $m$. This implies that $B_m$ exhibits a similar distribution across all $m$, leading to a loss of diversity in $H_p^m$. This loss is expressed as the correlation function between $B_m$ and $J_0(j_{0p}x)$, consequently reducing the reproducibility of $\delta f_p$. This also explains why the reproducibility of $\delta f$ in higher $v_\perp$ regions is consistently better than in lower $v_\perp$ regions. Figure 2 shows $B_{m=4}$ for $v_\perp^{ub} = 0.01c$ (black solid curve) and $c$ (red dotted curve). The values of $B_{m=4}$ are localized at $x \approx 0.7-1.0$ in both cases. Physically, this indicates that the electron velocity must be relatively large for the effect of the finite Larmor radius to become considerable.

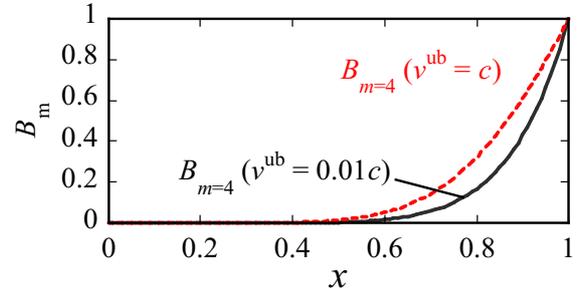

Figure 2. $B_{m=4}$ for $v_\perp^{ub} = 0.01c$ (black solid curve) and $c$ (red dotted curve). In both cases, $B_{m=4}$ values are localized at $x \approx 0.7-1.0$.

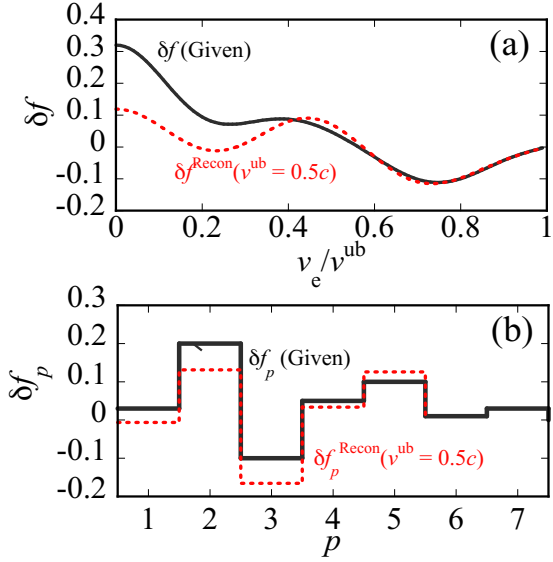

Figure 3. (a) The given $\delta f$ (black solid lines) and the reconstructed profiles for $v^{ub} = 0.5c$ (red dotted lines). The abscissa denotes the normalized electron velocity $v_\perp/v^{ub}$. (b) The corresponding $\delta f_p$, with abscissa representng the mode number in $p$-space. In this test, the harmonic ECE emissivity is given from $m = 2$–$5$, whereas the given $\delta f$ contains $p$-components from 1 to 7, specifically $\delta f_p = (0.03, 0.2, -0.1, 0.05, 0.1, 0.01,$ and $0.03)$.

The second test focuses on $\delta f$ reconstruction under ill-posed conditions, specifically when the wave number of harmonic ECE is less than $p_{max}$. This scenario is where the MEM–HT method truly shines. Figure 3 displays (a) the given $\delta f$ (black solid lines) and reconstructed profiles, and (b) the corresponding $\delta f_p$ at $v^{ub} = 0.5c$ (red dotted lines). In this test, the harmonic ECE emissivity is provided for $m = 2$–$5$, whereas the given $\delta f$ includes $p$-components ranging from 1 to 7, specifically $\delta f_p = (0.03, 0.2, -0.1, 0.05, 0.1, 0.01,$ and $0.03)$. Similar to the previous results shown in figure 1, the reproducibility of the $v_\perp$ profile appears less effective compared to that of the $p$-profile. Despite the lack of fundamental ECE measurements, the reconstructed $p$-profile closely matches the original with minimal error. This result further suggests that when the $\delta f_p$ values outside of $p_{max}$ are negligible relative to the $\delta f_p$ components for $p \leqq p_{max}$, the original $p$-profile can be effectively reconstructed.

## 5. Discussion and Summary

One of the optimal configurations for applying the MEM–HT method is an outboard ECE measurement in an optically thin, large aspecto ratio tokamak where the gradient of the toroidal magnetic field is sufficiently weak to prevent the overlap of harmonics of the electron cyclotron frequency. Furthermore, if the evanescent region between the upper-hybrid resonance and the right-hand cutoff frequencies at the edge is smaller than the vacuum wavelength of the X-mode for the fundamental ECE, measuring the fundamental ECE may be possible. In the MEM–HT method, because the effect of the propagator $\mathfrak{P}_m$ is eliminated, allowing essential information derived from the fundamental ECE component to be extracted without disturbance from propagation through the evanescent region. This approach enables the acquisition of the entropy distribution in phase space if it is combined with measurements in $\boldsymbol{k}$-space, e.g., using correlation techniques.

The MEM–HT method is a novel approach for reconstructing the fluctuation components of the EVDF $\delta f(v)$ and the associated electron entropy from the spectra of ECE harmonics in optically thin plasmas. Ths method utilizes the MEM in velocity space via the HT. Numerical tests confirm its validity. This method is applicable to various magnetized plasma situations, including both non-relativistic and relativistic electrons, except in cases of harmonic overlap or optically thick conditions. Furthermore, the method does not rely on ambiguous assumptions or radiometer calibration for measuring the EVDF. The reproducibility of both $\delta f$ and $\delta f_p$ improves significantly with increasing $v^{ub}$, indicating that this method is more effective when applied to higher electron energy ranges. However, even at $v^{ub} = 0.01c$, as demonstrated in the example presented in this article, acceptable reconstruction accuracy is still achieved. The measurement of relativistic frequency shifts can also be integrated with the proposed method to obtain the dependence of $\delta f(\boldsymbol{u})$ on the amplitude of the relativistivc parallel velocity $|u_{//}|$.

## Acknowledgements

The study was supported by grants from the Grants-in-Aid NSTC 113-2112-M-006 -017 and NSTC 113-2119-M-042A-001 from the National Science and Technology Council, Taiwan.